\documentclass[prb,aps,twocolumn,superscriptaddress,floatfix,citeautoscript,showpacs]{revtex4-1}
\usepackage{graphicx,rotating,subfigure,amsmath,amsfonts,amssymb,delarray,color}
\usepackage{dsfont}
\usepackage[T1]{fontenc}
\usepackage{multirow}
\usepackage{comment}

\newcommand{\tr}{\text{tr}}

\def\12{\frac{1}{2}}

\graphicspath{{figs/}}

\begin{document}
\bibliographystyle{apsrev}

\title{Logarithmic entanglement growth in two-dimensional disordered fermionic systems}

\author{Yang Zhao}
\affiliation{Shanxi Key Laboratory of Condensed Matter Structures and Properties, School of Science, Northwestern Polytechnical University, Xi'an 710072, China}
\author{Jesko Sirker}
\affiliation{Department of Physics and Astronomy, University of Manitoba, Winnipeg R3T 2N2, Canada}

\date{\today}

\begin{abstract}
We investigate the growth of the entanglement entropy
$S_{\textrm{ent}}$ following global quenches in two-dimensional free
fermion models with potential and bond disorder. For the potential
disorder case we show that an intermediate weak localization regime
exists in which $S_{\textrm{ent}}(t)$ grows logarithmically in time
$t$ before Anderson localization sets in. For the case of binary bond
disorder near the percolation transition we find additive logarithmic
corrections to area and volume laws as well as a scaling at long times
which is consistent with an infinite randomness fixed point. 
\end{abstract}

\maketitle

\section{Introduction}
Lately, localization phenomena in strongly correlated quantum
many-body systems have attracted considerable interest. It has been
discovered, in particular, that one of the hallmarks of {\it many-body
localization} (MBL) in one dimension is the logarithmic growth of the
entanglement entropy after quenching a system from a product state
\cite{ProsenZnidaric,BardarsonPollmann,LukinGreinerMBL,AndraschkoEnssSirker,EnssAndraschkoSirker}. 
It has also been studied---both theoretically and experimentally---how
localization in a many-body system can prevent information encoded in
the initial state from being completely erased as is expected in a
thermalizing system
\cite{AndraschkoEnssSirker,EnssAndraschkoSirker,BlochMBL,SchneiderBloch2,Bloch2dMBL,SirkerQuasiMBL}.

A natural question then is how such phenomena generalize to higher
dimensions. Experimentally, coupled chains of interacting fermions
with identical disorder have been investigated \cite{SchneiderBloch2} and
localization has been found to survive. The setup, however, is
fine-tuned and the dynamics starting from the chosen initial state
remains essentially purely one dimensional \cite{ZhaoAhmedSirker}. In a
later experimental study on a two-dimensional system with different
quasi-periodic potentials in {\it both} directions, indications for a
slowing down of the dynamics and a possible two-dimensional MBL phase
have been found \cite{Bloch2dMBL}. 

Theoretically, it remains an extremely difficult task to study large
interacting two-dimensional many-body systems with disorder in a
reliable manner. In this study we therefore want to take a step back
and investigate the dynamics after a quench in disordered
two-dimensional free fermion systems. Apart from being an interesting
problem in its own right, our study might also help in developing
criteria to identify possible two-dimensional MBL phases. We will
focus, in particular, on describing the growth of the entanglement and
the time evolution of local order parameters after a global quench
from a product state.

The entanglement entropy $S_{\textrm{ent}}$ of eigenstates of a
quantum system is, in general, expected to follow a volume law. Ground
states and low-lying excited states are, however, often an exception
and show instead pure area-law entanglement are an area law with
logarithmic corrections. Well understood are 1+1 dimensional quantum
field theories which show $S_{\textrm{ent}}\sim\ln(\xi)$ entanglement
in the ground state in the massive case with $\xi$ being the
correlation length. For critical systems, on the other hand,
$S_{\textrm{ent}}\sim\ln(\ell)$ where $\ell$ is the length of the
subsystem \cite{CalabreseCardy}. For $d$-dimensional gapless fermion
systems it has been shown, furthermore, that the ground state
entanglement scales as $S_{\textrm{ent}}\sim \ell^{d-1}\ln\ell$
\cite{Wolf_EntFermions,GioevKlich}.

\begin{figure}
	\centering 
	\includegraphics[width=0.9\columnwidth,angle=0]{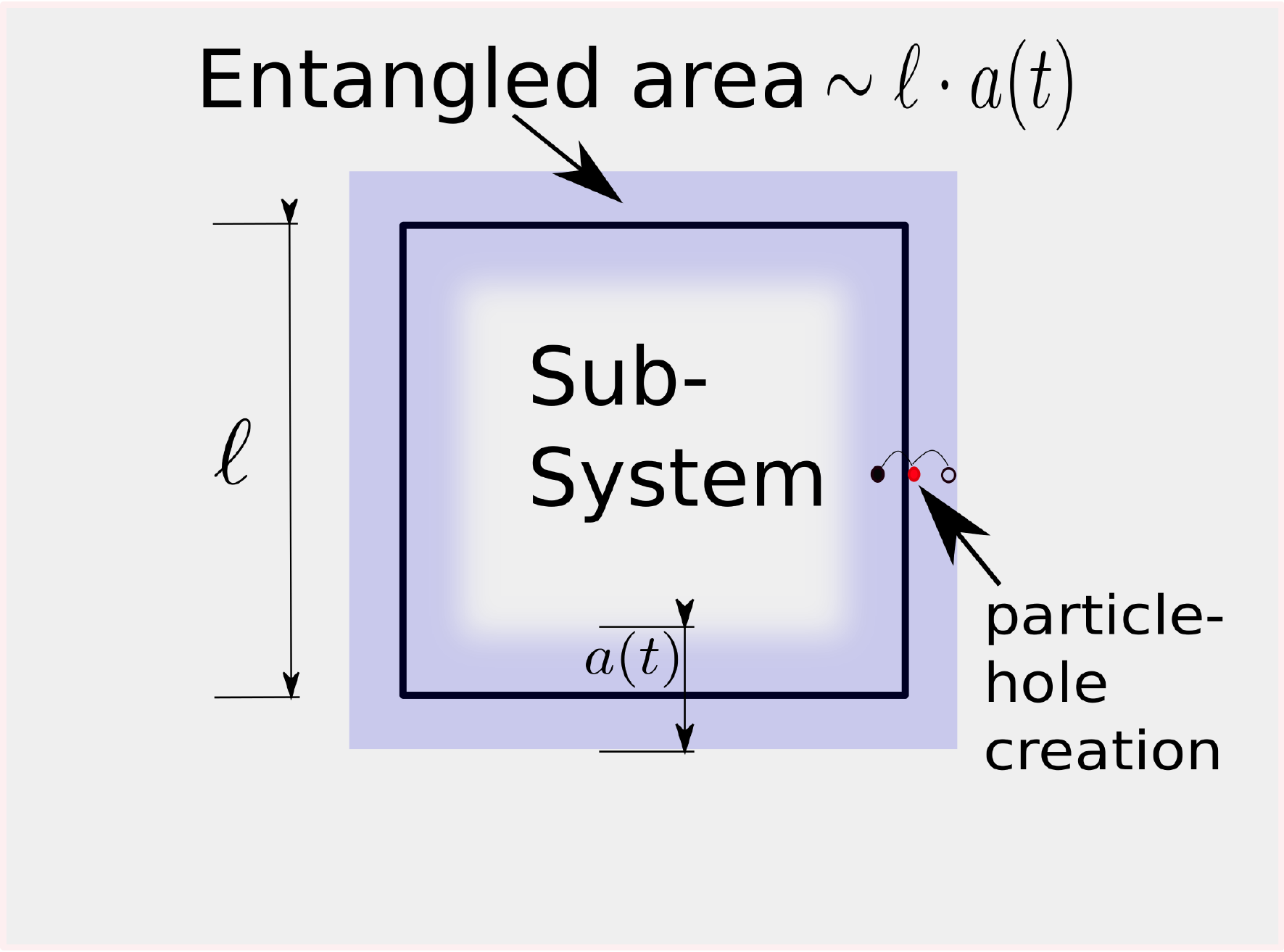}
	\caption{A quench in a two-dimensional system produces
	particle-hole pairs which propagate and create an entangled
	region which is proportional to the surface area $\sim \ell$
	of the subsystem times the length $a(t)$ over which the pairs
	have spread. For a clean system, in particular, a linear
	growth in time, $S_{\textrm{ent}}\sim\ell\cdot vt$, with
	velocity $v$ is thus expected for $vt\ll\ell$.}
\label{Fig1}
\end{figure}

A useful approach to understand the entanglement dynamics after a
quench is the quasi-particle picture \cite{CalabreseCardy2005}. A
quench produces particle-hole pairs which propagate with a velocity
$v$ in a clean system. The entanglement is then proportional to the
entangled region along the boundary between the subsystem and the rest
of the system. This entangled region is given by the area at time $t$
such that the particle (hole) of an excitation resides inside the
subsystem while the corresponding hole (particle) is outside, see
Fig.~\ref{Fig1}.
According to this picture one expects that for times $vt\ll \ell$ the
entanglement entropy grows as
\begin{equation}
\label{Eq1}
S_{\textrm{ent}} \sim \ell^{d-1}\cdot t 
\end{equation}
while $S_{\textrm{ent}}\sim \ell^d$ at times $vt\gg \ell$. For
quenches in free scalar field theories this picture is largely
confirmed also in two and three dimensions \cite{CotlerHertzberg}.

\section{Fermions with potential disorder}
In a classical system with potential disorder, fluctuations in the
density will typically relax in a diffusive manner. To obtain the
proper quantum mechanical picture for the spreading of excitations
after the quench for fermions with disorder, the disorder averaged
particle-hole propagator
\begin{equation}
\label{Eq2}
\mathcal{D}(r,t)=\langle\langle
\Psi^\dagger(r,t)\Psi(0,0)\rangle\langle\Psi(r,t)\Psi^\dagger(0,0)\rangle\rangle_{\textrm{dis}}
\end{equation} 
has to be calculated \cite{VollhardtWoelfle,AltlandSimons}. Apart from
a classical contribution where hole and particle propagate in the same
direction along the same path in configuration space (the diffuson
contribution), there are now also processes where particle and hole
traverse a path with opposite momenta (the cooperon contribution)
which will lead to interference effects. Ultimately, such quantum
corrections will result in a complete breakdown of diffusion---the
fermions become Anderson localized \cite{Anderson58}. It is well known
that in one and two dimensions any amount of disorder is sufficient to
completely localize all states
\cite{AbrahamsAnderson}. For the entanglement entropy at sufficiently 
long times after the quench we therefore expect an area law
$S_{\textrm{ent}}\sim
\ell^{d-1}\xi_{\textrm{loc}}$ if 
$\xi_{\textrm{loc}}\ll\ell$ where $\xi_{\textrm{loc}}$ is the
localization length, and $a(t\to\infty)\to\xi_{\textrm{loc}}$ see
Fig.~\ref{Fig1}.

Here we want to investigate how a disordered system evolves towards
this long-time limit. Contrary to the one-dimensional case where the
localization length $\xi_{\textrm{loc}}$ is of the order of the mean
free path $l_m$ and an initial ballistic spreading is immediately
followed by saturation with no room for diffusion, the situation is
much more complex in the two-dimensional case. Here the localization
length for weak disorder is exponentially large as compared to $l_m$
so that we might expect three distinct time regimes for the
entanglement entropy $S_{\textrm{ent}}(t)$. These regimes can be
characterized using the elastic scattering time $\tau=l_m/v$: (i)
$t\ll\tau$, initial ballistic increase, (ii)
$\tau\ll t \ll\tau\exp(\varepsilon\tau/\hbar )$, intermediate regime, and
(iii) $t\gg\tau\exp(\varepsilon\tau/\hbar)$, saturation (with
$\varepsilon$ being the characteristic energy scale of the
model). Understanding the intermediate regime is the main purpose of
this section.

To be specific, we will consider a square $L\times L$ lattice with
Hamiltonian
\begin{equation}
\label{Ham}
  H=-\sum_{\langle i,j\rangle}J_{ij}c^{\dagger}_{i}c_{j}+\sum_{i}D_{i}c^{\dagger}_{i}c_{i},
\end{equation}
where, $J_{ij}$ is the hopping amplitude between neighbouring sites
(we set $J_{ij}=J$ in this section), $c_{i}$ and $c^{\dagger}_{i}$ are
annihilation and creation operators of spinless fermions on site $i$,
and $D_{i}$ is the on-site disorder potential. The potentials $D_{i}$
are drawn randomly from a box $[0,D)$. Motivated by recent experiments
on cold atomic gases
\cite{BlochMBL} we choose as initial state a charge
density wave (CDW) configuration where singly occupied and empty sites
alternate in both spatial directions
\begin{equation}
\label{CDW}
|\Psi(0)\rangle = \prod_{i,j=1}^{L/2} c^\dagger_{2i,2j}c^\dagger_{2i-1,2j-1}|0\rangle \, .
\end{equation}
Here $|0\rangle$ denotes the vacuum state. We want to emphasize that
the results presented in the following do not qualitatively depend on
the specific initial state chosen as long as the setup is not
fine-tuned, e.g. in a way that the dynamics decouples and becomes one
dimensional \cite{ZhaoAhmedSirker}.

\subsection{Entanglement entropy}
To calculate the entanglement entropy we always choose a square with
size $\frac{L}{2}\times\frac{L}{2}$ in the middle of the $L\times L$
lattice as the subsystem, i.e.~$\ell\equiv L/2$ in the following. For
a non-interacting system, the entanglement entropy of the subsystem
can be obtained from its single-particle correlation matrix. If
$\zeta_i$ are the eigenvalues of the correlation matrix of the
subsystem then the entanglement entropy is given
by\cite{ChungPeschel,Peschel2004,PeschelEisler}
\begin{eqnarray}
\label{Sent}
S_{\textrm{ent}}(t)&\equiv& -\tr\{\rho_{\textrm{red}}\ln\rho_{\textrm{red}}\} \\
&=&-\sum_i \{\zeta_i\ln\zeta_i+(1-\zeta_i)\ln(1-\zeta_i)\} \nonumber \, ,
\end{eqnarray}
where $\rho_{\textrm{red}}$ is the reduced density matrix. Using exact
diagonalization (ED) we are thus able to investigate the dynamics in
relatively large two-dimensional lattices. We calculate disorder
averages using $\sim 1000$ samples for $L\leq 30$ sites and at least
$100$ samples for larger system sizes.

\begin{figure}
	\centering 
	\includegraphics[width=0.99\columnwidth,angle=0]{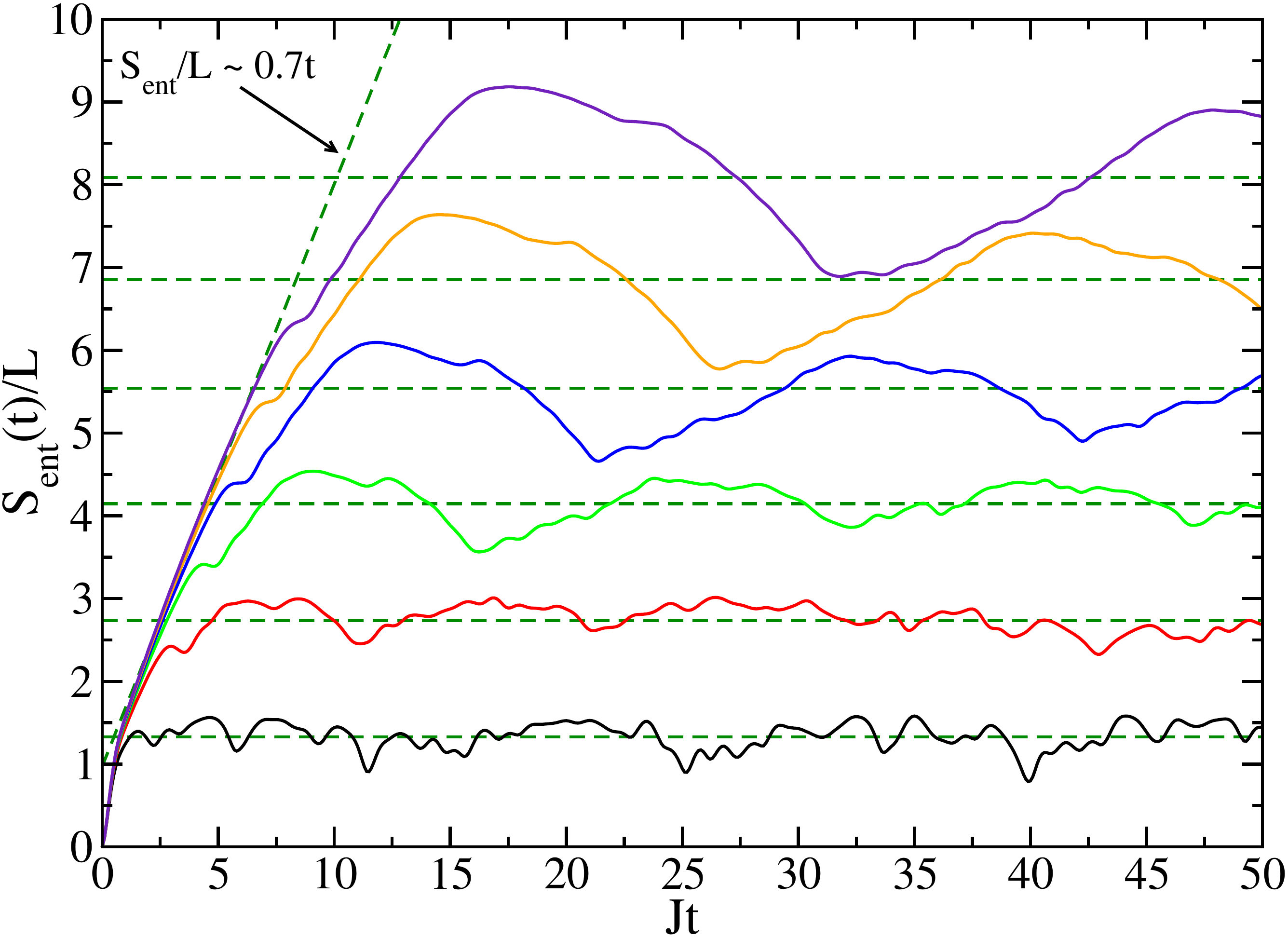} 
\caption{ED results for the entanglement entropy
	during unitary time evolution with the Hamiltonian \eqref{Ham}
	without disorder ($D_i\equiv 0$) starting from the CDW state
	\eqref{CDW}. Shown are results for system sizes $10\times 10$,
	$\cdots$, $60\times 60$ (solid lines, from bottom to top). The
	dashed lines are fits of the initial linear increase and the
	long-time asymptotics, respectively.}
\label{Fig2}
\end{figure}

As a first check, we consider the unitary dynamics caused by the
Hamiltonian
\eqref{Ham} without disorder, see Fig.~\ref{Fig2}.
After a quick initial increase at times $Jt\lesssim 1$ a linear
scaling sets in which lasts up to $t_0=L/4v=L/8$. Consistent with the
picture in Fig.~\ref{Fig1} the whole interior at this point has built
up some entanglement with the exterior and the increase of
entanglement slows down. At long times, the entanglement entropy
oscillates around $S_{\textrm{ent}}\approx 0.137 L^2$ consistent with
the expected volume law.

\begin{figure}
	\centering 
	\includegraphics[width=0.99\columnwidth,angle=0]{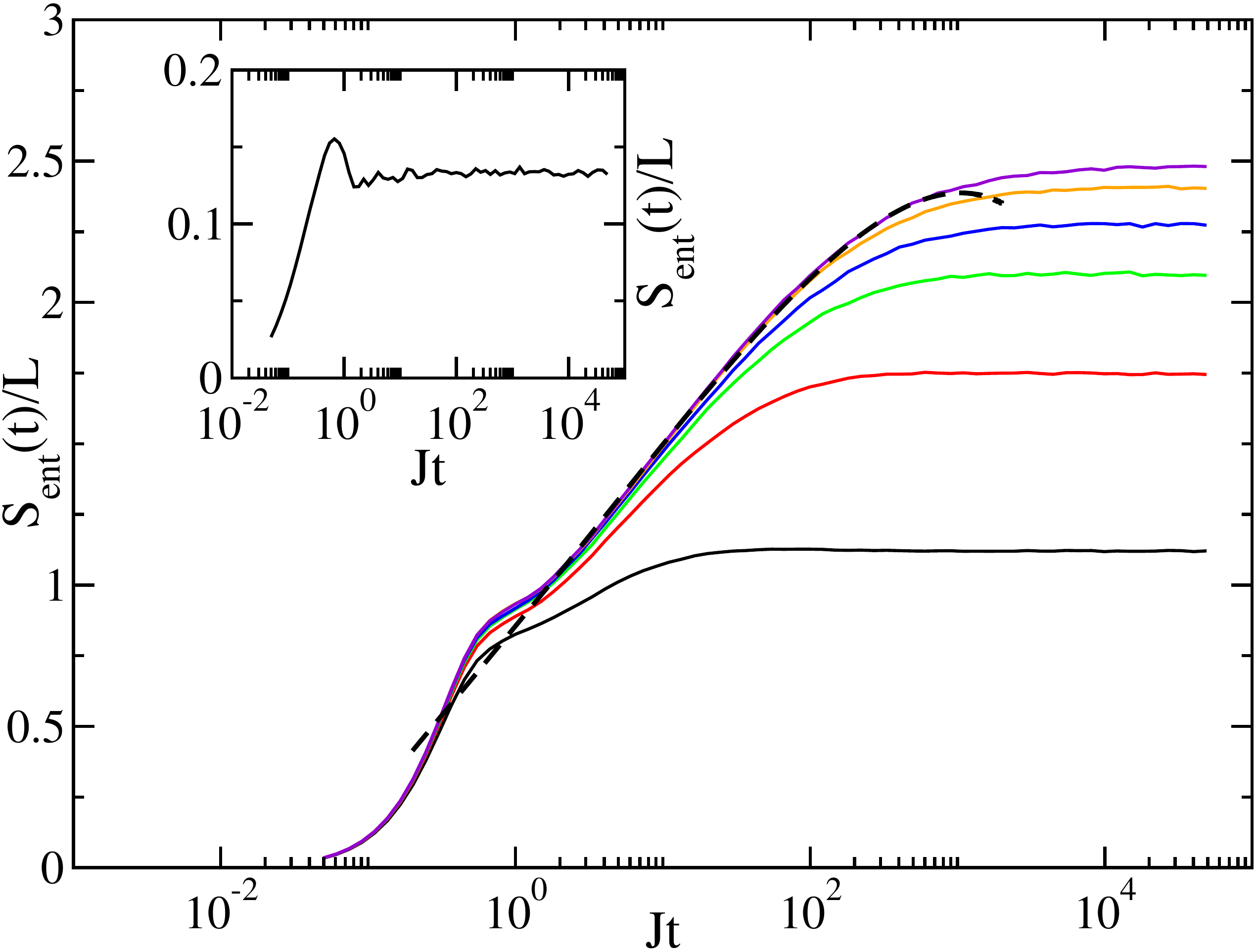}
	\caption{Inset: $S_{\textrm{ent}}(t)/L$ for $D=100$ and
	$L=40$. Main: Entanglement entropy for $D=10$ and system sizes
	$L=10,20,\cdots,60$ (from bottom to top). The dashed line is a
	fit in the intermediate time regime, see Eq.~\eqref{Sfit}.}
\label{Fig3}
\end{figure}

Next, we consider the case of strong potential disorder. Then an
initial increase of the entanglement entropy is immediately followed
by a saturation, see the inset of Fig.~\ref{Fig3}.
Most interesting is the case of small disorder where the localization
length $\xi_{\textrm{loc}}$ is much larger than the mean free path
$l_m$, see the main panel of Fig.~\ref{Fig3}. We can now indeed
identify three distinct time regimes: For times $Jt\lesssim 1$ there
is a fast initial increase. This is, however, now followed by an
extended intermediate time regime. Note also that with increasing
system size all data for $S_{\textrm{ent}}(t)/L$ fall onto a single
curve showing that the system obeys an area instead of a volume law
scaling. In the intermediate time regime---which for the chosen
disorder extends from $1\lesssim Jt
\lesssim 10^3$---the scaling is approximately logarithmic as in the 
one-dimensional many-body localized regime. In contrast to the latter
case, however, no interactions are present and the logarithmic scaling
is followed by a saturation at long times.

We show in the following that the logarithmic scaling can be
understood as a weak localization effect. Calculating the disorder
averaged particle-hole propagator
\eqref{Eq2} diagrammatically one finds
\begin{equation}
\label{Dqw}
\mathcal{D}(q,\omega)\sim \frac{iD_0q^2}{\omega+iD_0q^2}
\end{equation}
from summing up the ladder diagrams with non-crossing impurity lines,
the diffuson contribution.\cite{VollhardtWoelfle,AltlandSimons} The
particle-hole propagator shows diffusion at this level of
approximation with $D_0$ being the diffusion constant. Quantum
corrections to this result come predominantly from maximally crossed
diagrams, the cooperon contribution. These corrections involve
integrations over the diffusion pole which leads to logarithmic
corrections in two dimensions
\begin{equation}
\label{DiffPole}
\int \frac{q\,dq}{\omega+iD_0q^2} \stackrel{q\to 0}{\sim} \ln\omega \, .
\end{equation}
Here a cutoff for large $q$ implied. These corrections can be summed
up as ladder-type diagrams leading to a logarithmic correction of the
diffusion constant $D(\omega)$.\cite{GorkovLarkin} For the diffusive
spreading in the weak-localization regime this means that the mean
squared of the distance over which the particle-hole propagator has
spread is given by $\langle r^2(t)\rangle = 4D(t)t$ with a
time-dependent diffusion
constant\cite{NakhmedovPrigodin,SebbahSornette}
\begin{equation}
\label{Dt}
D(t) = D_0[1-(A\hbar/\varepsilon\tau)\ln(t/\tau)] \, 
\end{equation}
with some dimensionless constant $A$. This perturbative result is
expected to be valid for $\tau\ll
t\ll\tau\exp(\varepsilon\tau/(A\hbar)$. For the growth of the
entanglement entropy in this intermediate time regime this implies a
scaling 
\begin{equation}
\label{Sfit}
S_{\textrm{ent}}(t)/L = \mbox{const} + at^\alpha\sqrt{1-b\ln(t/\tau)}
\end{equation}
with $\alpha=1/2$. Fits using Eq.~\eqref{Sfit} do indeed show a good
agreement with the numerical data although the best results are
obtained using a smaller exponent $\alpha\approx 0.2$, see
Fig.~\ref{Fig3}. If we expand the scaling function \eqref{Sfit} around
its inflection point we find, in particular, that
$S_{\textrm{ent}}(t)/L \sim \mbox{const} +\ln(t/\tau)$. In the
intermediate time regime the entanglement entropy does grow
logarithmically.

\subsection{CDW order parameter}
\begin{figure}
	\centering 
	\includegraphics[width=0.99\columnwidth,angle=0]{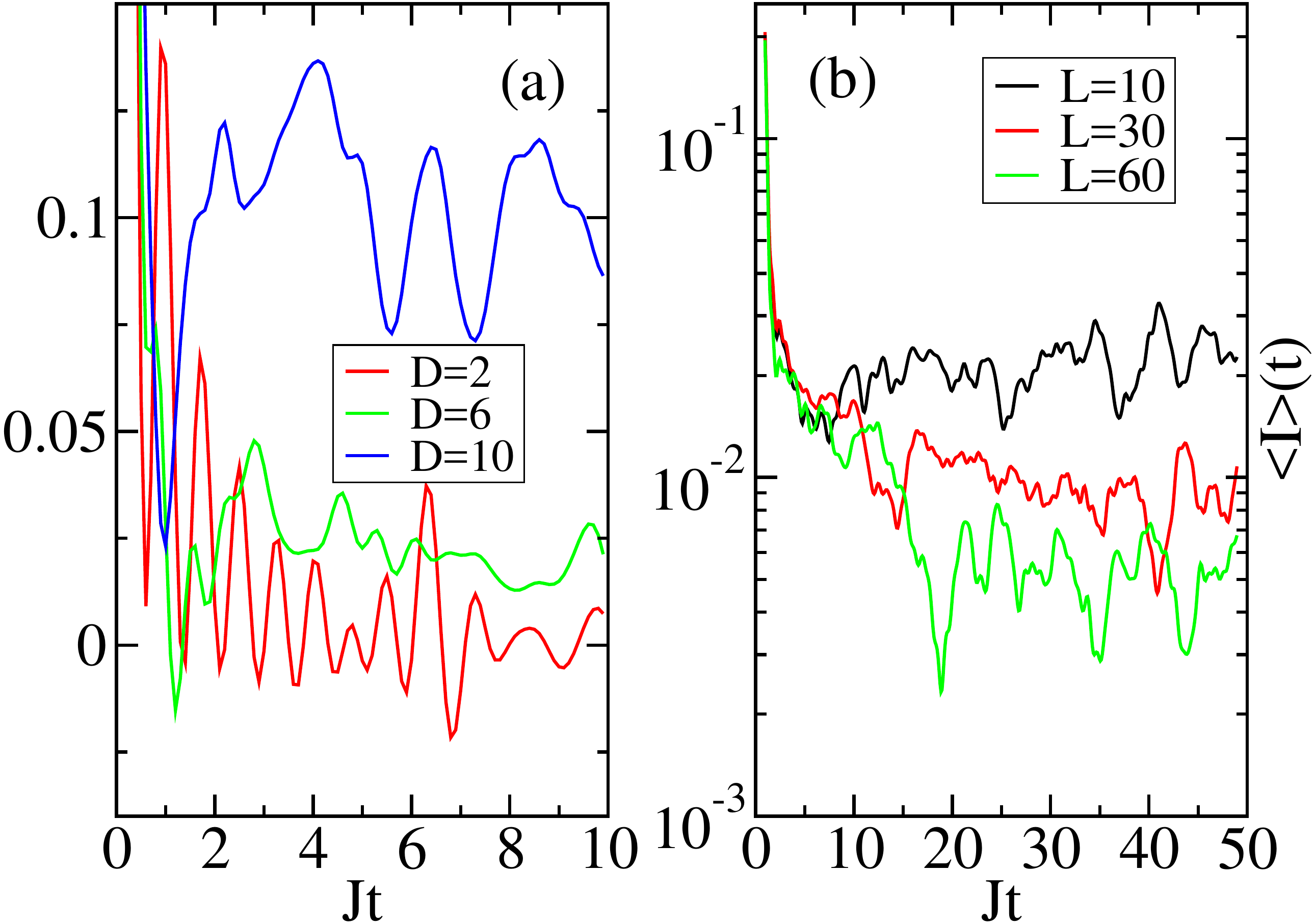}
	\caption{Order parameter $\langle I\rangle(t)$. Left: Data for
	a $20\times 20$ lattice with box potential disorder
	$D=2,6,10$. Right: Box potential disorder $D=5$ and lattice
	sizes $L=10,30,60$. For clarity, running averages over
	intervals $Jt=2$ are shown. After an initial decrease on the
	scale $Jt\sim 1$, $\langle I\rangle(t)$ starts oscillating
	around the long-time mean.}
\label{FigCDW}
\end{figure}

We start the quench from the initial CDW state \eqref{CDW}. This state
has an order parameter
\begin{equation}
\label{CDW2}
I=\frac{2}{L^2}\sum_{i,j} (-1)^{i+j} n_{i,j} = \frac{2}{L^2}\sum_{k_x,k_y} c^\dagger_{k_x,k_y} c_{k_x-\pi,k_y-\pi}
\end{equation}
with $\langle\Psi(0)|I|\Psi(0)\rangle=1$ whose evolution we want to
monitor as a function of time. The Fourier representation of the order
parameter in \eqref{CDW2} makes it clear that we are now looking at
the time evolution of a single-particle Green's function instead of
the particle-hole propagator \eqref{Eq2} relevant for the time
evolution of the entanglement entropy. This means, in particular, that
interference effects responsible for the weak localization phenomena
discussed in the previous section are expected to be absent. The data
shown in Fig.~\ref{FigCDW} are consistent with these expectations. A
well-defined intermediate time regime does not exist. Note that in the
numerics we calculate $\langle I\rangle(t)$ by taking the difference
in occupation of two neighboring sites in the middle of the square
lattice in order to reduce finite size effects.

For a clean two-dimensional square lattice it is easy to show that the
order parameter in the thermodynamic limit will completely decay as
$\langle I\rangle(t)=J^2_0(4Jt)\stackrel{t\to\infty}{\sim} 1/t$ where
$J_0$ is the Bessel function of the first
kind\cite{ZhaoAhmedSirker}. Fig.~\ref{FigCDW}(a) shows that once the
localization length becomes smaller than the system size, the order
parameter does no longer decay completely but rather oscillates around
a non-zero value after a quick initial decay on a time scale $t\sim
1/J$. While boundary effects remain present in the oscillations around
the mean value, Fig.~\ref{FigCDW}(b) shows that the long-time average
starts to converge when increasing the system size.

\subsection{Binary disorder and percolation threshold}
\begin{figure}
	\centering
	\includegraphics[width=0.99\columnwidth,angle=0]{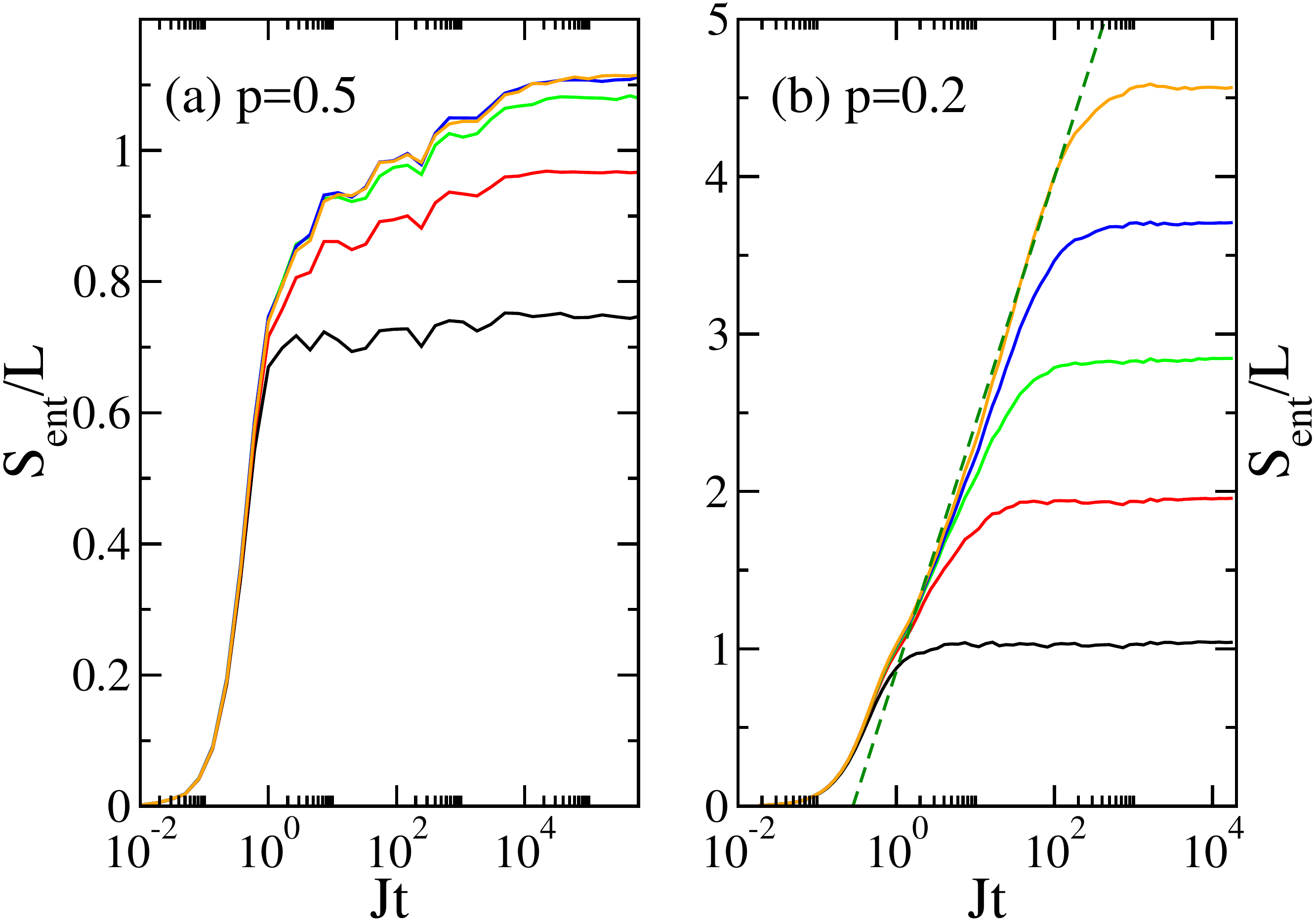}
	\caption{Binary potential disorder for $D=100$ and system
	sizes $L=10,20,\cdots,50$: (a) For $p=0.5$ the entanglement
	entropy shows area law scaling at long times. (b) Volume law
	scaling for $p=0.2$. The dashed line is a logarithmic fit of
	the intermediate time behavior.}
\label{Fig4}
\end{figure}

Another interesting question for two-dimensional systems with
potential disorder is the behavior near the percolation threshold. If
we consider strong binary disorder $D_i\in\{-D/2,D/2\},\, D\gg 1$ then
the system will be effectively cut into independent clusters of
potential $+D$ or $-D$ with particles unable to hop from one to the
other on time scales $t\ll
D/J^2$.\cite{AndraschkoEnssSirker,EnssAndraschkoSirker,SirkerQuasiMBL}
The site percolation threshold for a two-dimensional square lattice is
$p_c\approx 0.592746...$. I.e., if we have a probability $0.4\lesssim
p
\lesssim 0.6$ for a site to have potential $+D/2$ ($1-p$ for a site to
have potential $-D/2$) then both clusters will be
non-percolating. Otherwise, one of the clusters will be
percolating. In Fig.~\ref{Fig4} results for the entanglement entropy
for binary potential disorder $D=100$ and $p=0.5,\,0.2$, respectively,
are shown.
For $p=0.5$ we find the expected area law scaling, i.e., results for
$S_{\textrm{ent}}(t)/L$ do converge to a single curve for
$L\to\infty$. More interesting is the percolating case $p=0.2$. Apart
from the expected volume law scaling at long times, there is an
interesting intermediate time regime which is quite different from the
linear scaling due to the ballistic spreading of particle-hole pairs
in the clean case shown in Fig.~\ref{Fig1}. Instead, the time
dependence of $S_{\textrm{ent}}(t)$ seems to be consistent with a
logarithmic increase. This is somewhat surprising given that the
boundaries of a percolating cluster do perform a random walk and one
might thus at least classically expect a power-law dependence. Quantum
mechanically the problem is, of course, much more complicated. The
randomly shaped percolating clusters will result in many different
propagation paths which can interfere with each other. While this
might qualitatively explain why the entanglement growth is much slower
than in the clean case, we cannot offer a proper quantitative theory
at this point. Importantly, there seems to be a second
mechanism---apart from weak localization discussed previously---which
can lead to a logarithmic or almost logarithmic scaling of the
entanglement entropy at intermediate times. A logarithmic increase of
$S_{\textrm{ent}}(t)$ is therefore not a 'smoking gun' for many-body
localization as it is considered to be in the one-dimensional case.

\section{Fermions with bond disorder}
In contrast to potential disorder which always leads to localization
in one and two dimensions, bond disordered systems can display in
addition to a localized phase also infinite randomness fixed points
(IRFP) where the system is
critical.\cite{EggarterRiedinger,Fisher94,Fisher_random_Ising,Fisher_random_Ising2}
In one dimension, eigenstates show a $S_{\textrm{ent}}\sim\ln \ell$
scaling at an IRFP. Since length is expected to scale as $\ell\sim|\ln
t|^\Psi$ with dynamical critical exponent $\Psi$, the entanglement will
therefore grow extremely slowly as $S_{\textrm{ent}}(t)\sim\ln\ln(t)$
at long times. Numerically, such log-log scaling of the entanglement
entropy after a quench has been observed in the critical transverse
Ising chain\cite{IgloiSzatmari} and the XX-chain
\cite{ZhaoAndraschkoSirker}. The situation is less settled in higher
dimensions. For the critical two-dimensional transverse Ising model, an
area law with multiplicative logarithmic corrections,
$S_{\textrm{ent}}\sim \ell\ln(\ln \ell)$, has been suggested in
Ref.~\onlinecite{LinIgloiRieder} while a second numerical study,
Ref.~\onlinecite{YuSaleurHaas}, has interpreted their results in terms
of an additive logarithmic correction, $S_{\textrm{ent}}\sim a\ell+b\ln
\ell$. In the latter study, it has been argued that the presence of an
additive logarithmic correction is related to percolation. An additive
logarithmic correction was later confirmed in large-scale numerical
strong disorder renormalization group calculations and the dependence
of the size of the logarithmic correction on the shape of the
subsystem was studied \cite{IgloiRTIM2,IgloiPerc}.

\begin{figure}
	\centering
	\includegraphics[width=0.99\columnwidth,angle=0]{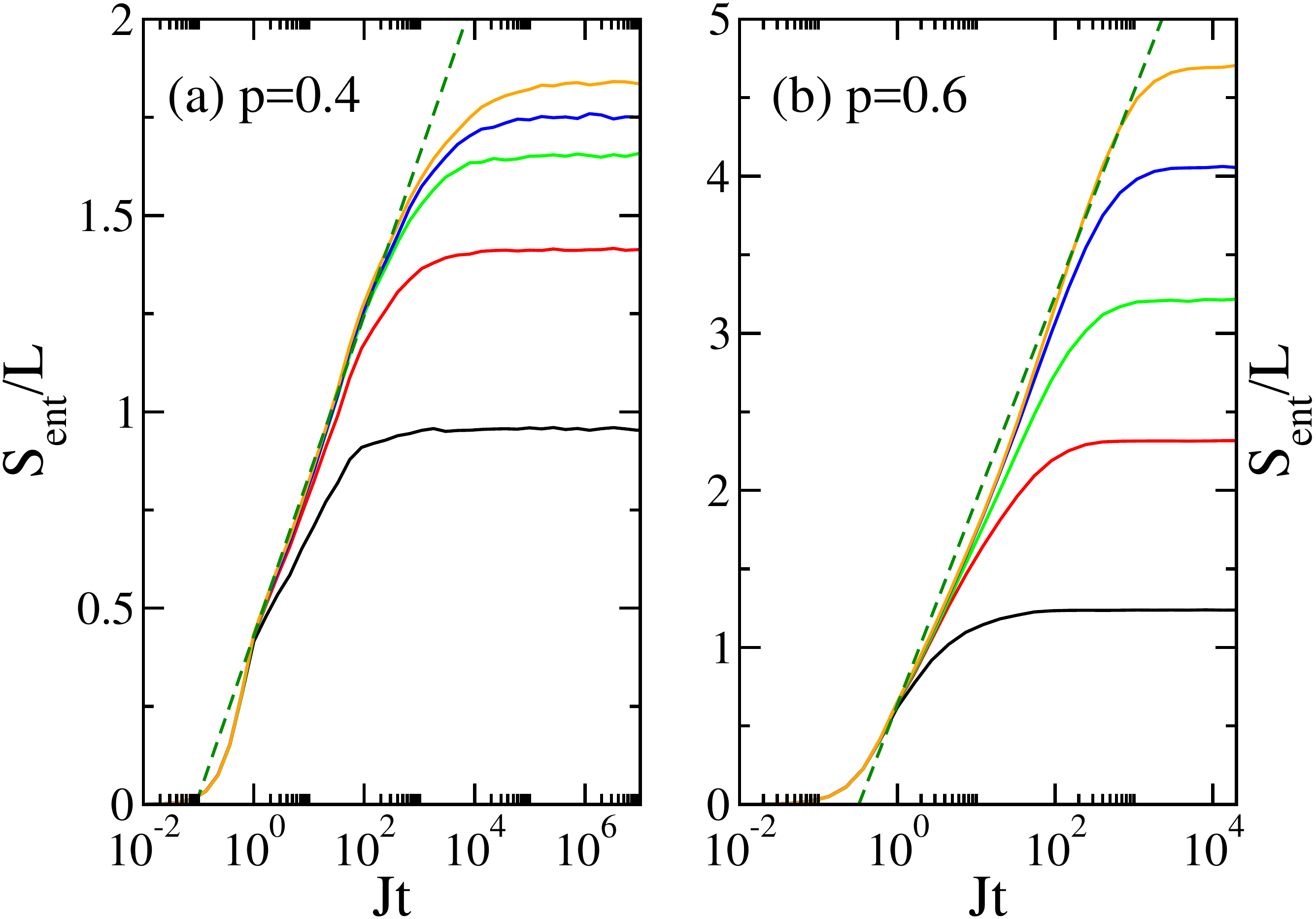}
	\caption{Binary bond disorder $J_{ij}\in\{0.05,0.95\}$: (a)
	for a probability $p=0.4$ to have a strong bond the long-time
	scaling seems to approximately follow and area law, while (b)
	for $p=0.6$ the scaling appears close to a volume law. The
	dashed lines are logarithmic fits at intermediate times.}
\label{FigPerc}
\end{figure}

\begin{figure}
	\centering
	\includegraphics[width=0.99\columnwidth,angle=0]{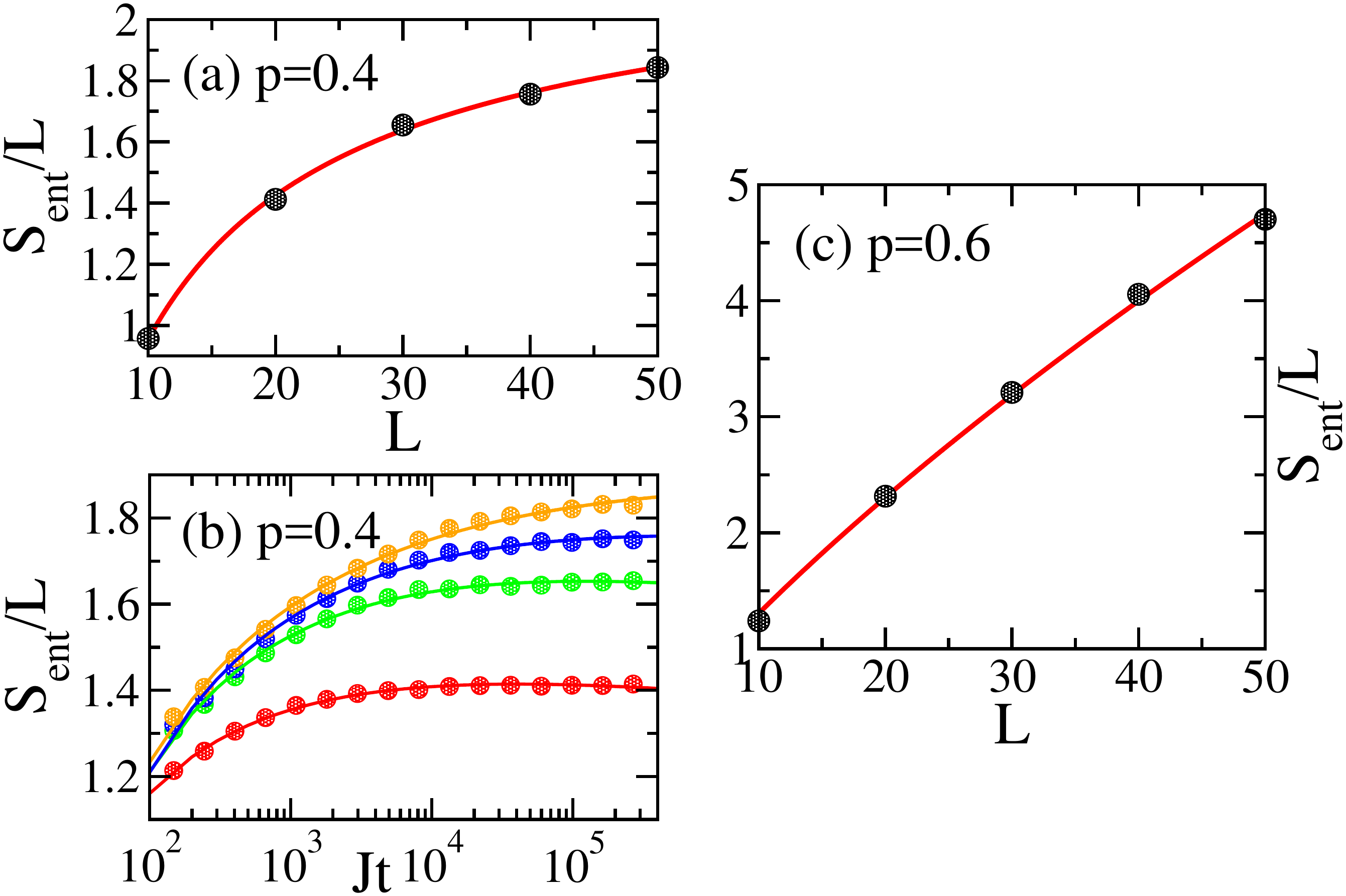}
	\caption{Binary bond disorder $J_{ij}\in\{0.05,0.95\}$: (a)
	Saturation values at long times versus $L$ (circles) and fit
	$S_{\textrm{ent}}/L = a + b(\ln L)/L$ (line). (b) Long-time
	behavior and fit $S_{\textrm{ent}}/L = a' + b'\ln (\ln t)/(\ln
	t)^\Psi$ with $\Psi\sim 0.4$. (c) Saturation values and fit
	$S_{\textrm{ent}}/L = \tilde a L + \tilde b L\ln L$.}
\label{FigPerc2}
\end{figure}

Here we want to examine the entanglement growth directly in the
microscopic two-dimensional free fermion model \eqref{Ham} with bond
disorder. In order to study the connection to percolation, we
concentrate on the case of binary bond disorder. If we have bonds
drawn from $J_{ij}\in\{0,1\}$ then the classical percolation threshold
is $p_c=0.5$. For this type of bond disorder we might thus expect that
the entanglement entropy follows an area law if the probability for a
bond being present ($J_{ij}=1$) is given by $p<p_c$ while a volume law
should hold for $p>p_c$. To study the generic dynamics near the
percolation threshold we present in Fig.~\ref{FigPerc} data for binary
disorder $J_{ij}\in\{0.05,0.95\}$ on a square $L\times L$ lattice with
a subsystem size $\ell=L/2$.
For both probabilities $p$ to have a strong bond shown in
Fig.~\ref{FigPerc}, the entanglement entropy per length $L$ increases
approximately logarithmically at intermediate times before starting to
saturate. While the scaling at long times in Fig.~\ref{FigPerc}(a) for
$p=0.4$ appears to be close to an area law, the curves for the largest
system sizes studied do not fall on top of each other and show a slow
convergence towards the saturation value. To investigate this behavior
further we show in Fig.~\ref{FigPerc2}(a) the saturation values at
long times for the case $p=0.4$ as a function of length $L$.
The entanglement per area is well fitted by $S_{\textrm{ent}}/L = a +
b(\ln L)/L$ supporting an area law with an additive logarithmic
corrections as suggested in
Ref.~\onlinecite{YuSaleurHaas}. Furthermore, the expected scaling
$L(t)\sim (\ln t)^\Psi$ at an IRFP yields a very good fit of the
long-time behavior, see Fig.~\ref{FigPerc2}(b). On the other side of
the percolation transition we find saturation values of the
entanglement entropy which are consistent with a volume law scaling
with an additive logarithmic correction, see
Fig.~\ref{FigPerc2}(c). Using again the expected scaling relation
between length and time leads to a logarithmic scaling consistent with
the data shown in Fig.~\ref{FigPerc}(b). Overall, our data are
consistent with additive logarithmic corrections to the area and
volume laws on either side of the percolation transition and IRFP
behavior, $L(t)\sim (\ln t)^\Psi$, at long times.

\section{Summary and conclusions}
Quenching two-dimensional free fermion systems with potential and
binary disorder, we have identified several cases where the
entanglement entropy $S_{\textrm{ent}}(t)$ shows a logarithmic growth
as a function of time $t$ as well as logarithmic corrections to area
and volume laws.

The behavior of $S_{\textrm{ent}}(t)$ for potential disorder can be
understood in a picture of particle-hole pairs created by quenching
the system. As is well known, the particle-hole propagator consists of
a classical diffuson and a quantum correction, the cooperon
contribution. In the intermediate time regime $\tau \ll t \ll
\tau\exp(\varepsilon\tau/\hbar)$,  where $\tau$ is the elastic scattering time, 
the interference effects due to the latter contribution lead to {\it
weak localization}. We have shown that weak localization results in a
logarithmic growth of the entanglement entropy per area,
$S_{\textrm{ent}}(t)/L
\sim \mbox{const} +\ln(t/\tau)$, before Anderson localization and 
thus a saturation of $S_{\textrm{ent}}(t)/L$ ultimately sets in at
times $t\gg\tau\exp(\varepsilon\tau/\hbar)$. We demonstrated,
furthermore, that the weak localization regime does not materialize in
a local order parameter: here the relevant quantity is the single
particle Green's function which---in contrast to the particle-hole
propagator---does not show interference effects. For experiments on
cold atomic gases this means that a monitoring of local order
parameters as in Ref.~\onlinecite{BlochMBL,Bloch2dMBL} would not be
sufficient to observe weak localization. Instead, the recently
demonstrated direct measurements of number and configurational
entropies\cite{LukinGreinerMBL} offer an avenue to explore this regime
if generalized and applied to the two-dimensional case.

For the case of bond disorder, we have concentrated on a binary
distribution where strong bonds occur with probability $p$ and weak
bonds with probability $1-p$. For strong binary bond disorder, we have
found qualitatively different behavior for the entanglement entropy
after quenching from a product state based on whether or not $p$ is
smaller or larger than the classical percolation threshold
$p_c=0.5$. While $S_{\textrm{ent}}(t)$ at long times is approximately
showing an area law scaling for $p<p_c$ we find approximately a volume
law for $p>p_c$. Interestingly, there are additive logarithmic
corrections present in both cases consistent with IRFP behavior. For
the case $p<p_c$ the logarithmic corrections lead, in particular, to a
very slow increase of the entanglement entropy at long times,
$S_{\textrm{ent}}(t)/L\sim \ln\ln t/(\ln t)^\Psi$, with dynamical
critical exponent $\Psi$.

In conclusion, we have demonstrated that the entanglement entropy of
two-dimensional disordered fermion systems shows a very rich and
interesting behavior as a function of time after a global quench. We
have explained the logarithmic growth in the potential disorder case
by weak localization physics and the logarithmic corrections to area
and volume laws as well as the long-time scaling in the bond
disordered case by IRFP physics. While our work does not address the
interacting case and the question whether or not many-body localized
phases exist in two-dimensional quantum systems, the results obtained
here might be helpful to develop criteria to distinguish possible MBL
phases from Anderson physics and IRFP behavior.

\acknowledgments
Y. Zhao acknowledges support by Fundamental Research Funds for the
Central Universities (3102017OQD074). J.~Sirker acknowledges support
by the NSERC Discovery grants program (Canada) and by the DFG via the
Research Unit FOR 2316 (Germany).


\end{document}